\documentclass{aa}
\usepackage{psfig} 

\def\mcg{MCG--6-30-15\/}
\def\xmm{{\sl XMM-Newton }}

\def\G{$\Gamma$}

\newcommand{\ltsima} {$\; \buildrel < \over \sim \;$}
\newcommand{\gtsima} {$\; \buildrel > \over \sim \;$}
\newcommand{\lta} {\lower.5ex\hbox{\ltsima}}
\newcommand{\gta} {\lower.5ex\hbox{\gtsima}}

\def\approxlt{\mathrel{\hbox{\rlap{\lower.55ex \hbox {$\sim$}}
        \kern-.3em \raise.4ex \hbox{$<$}}}}
\def\approxgt{\mathrel{\hbox{\rlap{\lower.55ex \hbox {$\sim$}}
        \kern-.3em \raise.4ex \hbox{$>$}}}}

\begin{document}

\title{X-ray to UV Variability Correlation in \mcg}  
\author{P. Ar\'evalo\inst{1} \and I.Papadakis\inst{2} \and B. Kuhlbrodt\inst{3,4} \and W. Brinkmann\inst{1}}
\offprints{P. Ar\'evalo, parevalo@mpe.mpg.de}
\institute{
Max-Planck-Institut f\"ur extraterrestrische Physik,
         Postfach 1312, D-85741 Garching, Germany
\and Department of Physics, University of Crete, P.O. Box  2208,71 003 Heraklion, Crete, Greece
\and Astrophysikalisches Intitut Potsdam, An der Sternwarte 16,14482 Potsdam, Germany
\and Hamburger Sternwarte, Gojenbergsweg 112, 21029 Hamburg, Germany
}
\date{Received 5 August 2004/Accepted 22 September 2004 }

\abstract{ We used a $\sim 300$ ks long \xmm\ observation of the Seyfert 1 galaxy \mcg\ to
study the correlation between the 0.2--10 keV
X-ray and the 3000--4000 \AA~$U$ bands. We found a significant
correlation peak at a time lag of $\tau_{\rm max}\sim 160$ ks where the UV
flux variations preceded the variations in the X-ray band. We
interpret this result as evidence in favour of Comptonisation models
where the observed X-rays are produced through Compton up-scattering
of thermal UV seed photons from an accretion disc, as this process
naturally predicts the UV variations to precede similar flux
variations in the X-rays. The length of the time lag favours models
where the observed UV and the seed-photon-emitting regions are
connected by perturbations of the accretion flow traveling inwards
through the disc, affecting first the main-$U$-band-emitting radii and then the
innermost region where the bulk of the seed photons is expected to be
produced. Finally, the absence of significant features in the correlation function with X-ray flux variations preceding those in the UV indicates that the observed $U$-band
photons are not mainly produced through reprocessing of hard X-rays in
this source.

}
 
\titlerunning{X-ray to UV Variability Correlation in \mcg}
\authorrunning{Ar\'evalo et al.}
\maketitle\section{Introduction} 
The spectral energy distributions of Active Galactic Nuclei (AGNs) are
extremely broad, often ranging from radio wavelengths up to gamma
rays. Though this is well established, the origin of the emission
is still a matter of debate. In general most of the AGN's luminosity
is in the so called Big Blue Bump in the optical-UV regime, and a considerable fraction of the energy is emitted in the X-ray
band. The study of the variability of the sources in these energy bands and
their correlations can provide important information on the emission
mechanisms.

The most widely accepted model for the energy release in AGN is
accretion of matter onto a supermassive black hole, and the accretion flow is thought
to form an optically thick disc. In the standard picture this accretion disc radiates thermally
mainly in the optical/UV band, for AGN black hole masses of $\sim 10^6 - 10^8$ M$_\odot$. The
production of X-rays is commonly attributed to Compton up-scattering
of UV-photons (e.g. Sunyaev \& Titarchuck \cite{Sun}) by hot
electrons in a corona (Haardt \& Maraschi \cite{Haardt}). Under this hypothesis the UV and X-ray light
curves are expected to be correlated with the X-rays lagging the
UV. However, it is also possible that the X-rays produced this way would illuminate the disc or other
surrounding optically thick material and produce UV radiation through reprocessing
(e.g. Guilbert \& Rees \cite{Gui}). If the bulk of the observed UV continuum
arises from this thermal re-emission the resulting light curves would
again be correlated but this time the UV should lag the X-rays. Some models where
the UV and X-ray emission come from a single continuum process predict
no time lags between the bands. In principle simultaneous observations
of these two energy bands with good time resolution and sampling
should provide information about the physical and geometrical
conditions of the emission region and help to discriminate between the
models.

Several multi-wavelength monitoring campaigns of AGN conducted
in the previous decade aimed to search for
correlations between X-rays and optical/UV emission. A definite
correlation could be found in only a few cases and the sign or existence
of a time lag differed from case to case (see Maoz et al. \cite{Maoz} for the
puzzling results of their long-duration campaign of NGC3516 together with a
summary of previous studies). The existence of positive as well as
negative time lags suggests that different processes could be
dominating the emission at different times and, in general, does not imply any
simple relation between the energy bands. Some cases where no lags were
found might be explained in terms of inappropriate sampling of the
light curves, if the time lags are for example much shorter that the
typical spacing between data points.

In this paper we study the temporal variations of the Seyfert1 galaxy \mcg\ ($m_{B}$$\sim$13.8, $z$=0.00775). This galaxy is famous for its broad and
skewed Fe K$_{\alpha}$ line (Tanaka et al. \cite{Ta}), which is consistent with
fluorescent emission from a disc close to the central supermassive
black hole. This spectral feature is strong evidence for the
accreting black hole model and limits the primary hard X-ray emitting
region to a few Schwarzschild radii $ R_{\rm S}$ (Fabian \& Vaughan \cite{Fab2}). \mcg\ has been
extensively observed in X-rays where it shows large and rapid flux
variability (see eg. Iwasawa et al. \cite{Iwa};
Fabian et al. \cite{Fab2}). Timing analyses indicate a featureless, red-noise-like power density spectrum with a possible break at $10^{-4}$ Hz as
its only characteristic time scale (Vaughan et al. \cite{Vaug1}).

The aim of the work presented here is to determine the relationship
between the UV and X-ray light curves of \mcg. The results of
cross-correlation analyses of the variability of the light curves are
compared with the predictions of different emission models to estimate
their applicability to this object. We make use of simultaneous X-ray
and UV data obtained with the instruments on board \xmm . The
observation spans approximately 430 ks (i.e. $\sim 5$ days) and
the time resolution, limited by the sampling time of the UV light
curves, is $\sim$ 1.2 ks. Therefore we can use these data to search
for correlations between the X-ray and UV light curves on time scales
of a few ks with lags up to a few days.

The paper is organized as follows: In Sect. 2 we describe the data reduction and the
construction of the light curves. Sect. 3 contains the cross
correlation analysis, implications of the results are discussed in Sect. 4
and we summarize our conclusions in Sect. 5.
\section{Data reduction and light curves} 
\mcg\ was observed with
XXM-Newton between July $31^{st}$ and August $5^{th}$ 2001, during
revolutions 301, 302 and 303. The observations spanned a total time of
$\sim 430$ ks with 80 ks, 122 ks and 123 ks stretches of
scientifically useful, uninterrupted exposures. For the analysis we
used data from the EPIC-PN detector (Str\"uder et al. \cite{Strue}) to
construct the X-ray light curve and from the Optical Monitor, OM (Mason et al. \cite{Ma}) for the UV.

The X-ray data used here have already been analyzed extensively by
several authors. Vaughan et al. (\cite{Vaug1}) studied in detail the
time variability of the X-ray continuum in the PN energy band. The
broad Fe K$_{\alpha}$ line in the spectrum has been studied and
modeled by Fabian et al. (\cite{Fab1}), Fabian \& Vaughan
(\cite{Fab2}), Miniutti et al. (\cite{Mi}) and Ballantyne et
al. (\cite{Ball}) using this same data set. Vaughan \& Fabian
(\cite{Vaug2}) tested different spectral models for the PN 3--10 keV
band, and the soft X-ray spectrum was studied by Turner et
al. (\cite{Tur1}, \cite{Tur2}) using the reflection grating
spectrometer (RGS) data of the same observation.

\subsection{X-ray light curves}
All EPIC-PN exposures were taken in the small-window (SW) mode using
the medium filter. We processed the data using XMMSAS version
5.4.1. The light curve was constructed following the procedure
described in Fabian et al. \cite{Fab1}, differing only in extraction
radius (30$\arcsec$ in the previous analysis, 50$\arcsec$ in the present work). This
bigger radius was chosen to include at least $ 90\% $ of the energy of
a point source (Ghizzardi \& Molendi \cite{Ghi}) and results in a
slightly higher count rate (6\% difference) than that presented in the work mentioned above.

The background-subtracted light curves were binned in different bin
sizes and the corresponding errors were calculated by propagating the
Poissonian noise. Bins with data gaps were corrected for their
effective exposure times. Since all exposures were taken in the same
observation mode we did not correct for mode-dependent exposure time
losses such as dead time or out-of-time events.

\begin{figure}[h]
\psfig{figure=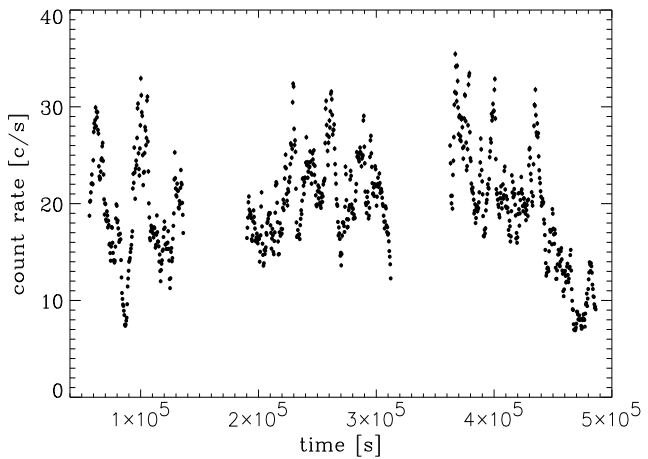,width=8.5cm,height=6.0cm,%
   bbllx=87pt,bblly=440pt,bburx=515pt,bbury=722pt,clip=}
\caption {X-ray light curve from the PN data in 400 s bins. Time is in seconds starting from 0:00:00 hrs of July $31^{st}$, 2001. The errors in the count rate are typically smaller than the symbols.}
\label{pntotal}
\end{figure}

In general the background activity was low, on average less than $1\%$
of the source count rate. Only in the last few ks of each exposure did the
background increase significantly. These segments of the light curves
were excluded from the analyses. 
The final light curve is shown in Fig. \ref{pntotal}.

The X-ray light curve displays large amplitude variations in all
observable time scales with an overall minimum-to-maximum ratio of
$\sim 5$. The power spectrum of this light curve has approximately a
broken power-law shape and the frequency of the break at $\sim
10^{-4}$ Hz indicates the presence of a characteristic time scale at
$\sim 10^4$ s in the system. The average luminosity of the source in
the 0.2--10.0 keV band is $\sim 10^{43}$ erg/s and the minimum-to-maximum 
luminosity difference has a value of $\Delta L \sim 1.5 \times
10^{43}$ erg/s (Vaughan et al. \cite{Vaug1}).

\subsection{UV light curves}
\begin{figure}[h]
\psfig{figure=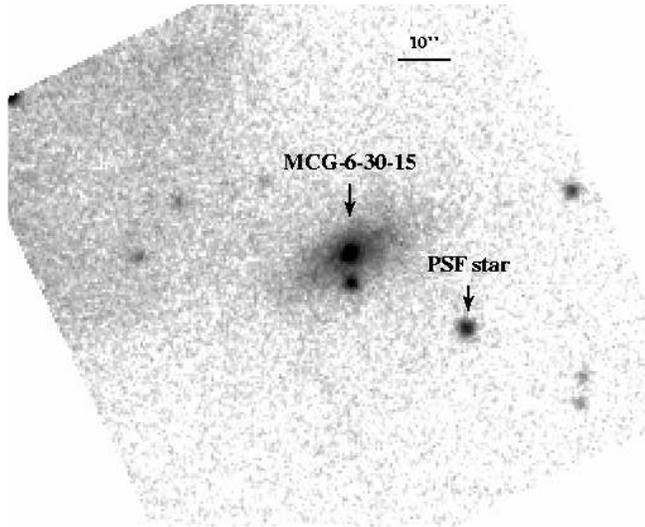,angle=0,width=8.5cm,height=7.0cm,%
   bbllx=158pt,bblly=307pt,bburx=438pt,bbury=535pt,clip=}
\caption {OM image of \mcg. The host galaxy is clearly seen around the brightest point source, which corresponds to the AGN. The secondary object 6'' to the south of the nucleus and the non-uniform background pattern were taken into account in the fitting procedure.}
\label{image}
\end{figure}

 The OM is a 30 cm optical/UV telescope co-aligned with the X-ray
telescopes on board \xmm . It is sensitive in the 1600-6000 \AA\
wavelength range and its typical point spread function has a full
width at half maximum of 2.2$\arcsec$ in the $U$ band. For this observation
the OM was operated in imaging mode using the $U$ filter 
(3000-4000 \AA ). We processed the data using the task {\sc
omichain} of XMMSAS version 5.4.1. The data consist of a sequence of
274 snapshot exposures 800 s long typically separated by time gaps of
320 s. Fig. \ref{image} shows one of the OM exposures on logarithmic
scale.  The $U$-band luminosity of the AGN is smaller than that of the
total host galaxy so the Poissonian error in the total
(galaxy+nucleus) counts introduces a significant scatter that may mask
the intrinsic variability of the nucleus. It was therefore essential
to separate the contribution of the nucleus from that of the rest of
the galaxy to obtain an accurate measure of the AGN variability. To do
this we used the algorithm developed by Kuhlbrodt et al. (\cite{Kuhl})
which decomposes the image of a galaxy into disc, bulge and/or
nuclear point source components.

The principle behind the algorithm is that the light
distribution of the nucleus is modeled with the image of a nearby star. Since
both the star and the active nuclear region are not resolved in the UV,
their images are representative of the point-spread function (PSF) of the optical
system.  Spatial or temporal variation of the PSF do not play a role
in our case: as a nearby star with high signal-to-noise was always
present in the images, variations in the PSF were taken into account by determining its actual shape in every exposure.  This star, labeled PSF star in Fig. \ref{image}, is only used for the PSF modeling and not for photometric comparison. The model of the galaxy consists of the
exponential disk profile by Freeman (\cite{Free}) and a de Vaucouleurs
(\cite{deV}) profile for the bulge, expanded to two dimensions with
elliptical isophotes (see Kuhlbrodt et al. \cite{Kuhl} for details) and convolved
with the PSF. The nuclear and galaxy components are then
fitted simultaneously.  

\subsubsection{Image decomposition}
As a first approach \mcg\ was fitted using a three-component model
accounting for a disc, a bulge and a nuclear point source. The bulge contribution to the galaxy flux is small, typically $\sim$ 10\% of the disc flux. This component was subsequently discarded from the fitting procedure to improve the accuracy in the determination of the nuclear flux. 

\begin{figure}[h]
\psfig{figure=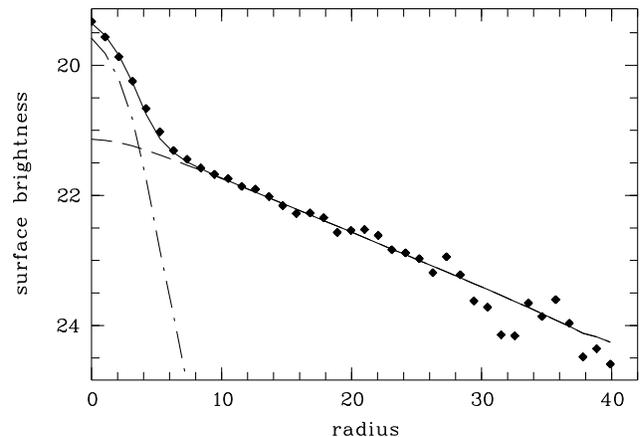,angle=270,width=8.5cm,height=5.8cm,%
   bbllx=99pt,bblly=37pt,bburx=298pt,bbury=330pt,clip=}
\caption {Light profile fit with a two-component model. The radius is in pixels (1 pixel=0.48'') and the surface brightness in arbitrary units. The solid line shows the total model fit to the data (filled diamonds). The dashed line shows the disc contribution, and the dashed-dotted line the nuclear component.}
\label{disc}
\end{figure}

As can be seen in Fig. \ref{disc} the two-component (disc+nuclear point source)model is a good
approximation and fits the light profile accurately. This figure
shows the model fit, where the dashed line represents the disc, the
dashed-dotted line the nuclear component and the solid line represents
the sum of the components fitted to the data points (filled diamonds).
It is important to note that although \mcg\ has been classified as an
E/S0 galaxy based on the galaxy morphology as seen in the optical
band, in the $U$ band it is very well fitted by a Freeman disc model
only.

We used the two-component model to fit all 274 images individually,
allowing only for variable nucleus and disc normalisations. The disc
fluxes obtained showed a very small scatter in their values, entirely
accountable for by Poissonian noise, and gave no indication of strong
systematic changes in the photometric performance of the
detector. However, as the nuclear flux is small compared to the disk
flux, $F_{\rm nuc}/F_{\rm disc} \sim 0.17$, and the simultaneous
determination of both fluxes is slightly degenerate, even small errors
in the disc flux could result in significant artificial variations in
the nuclear light curve. To make sure that the nuclear variability
present in the light curve was not an artifact of this degeneracy we
modified the routine in order to enforce a constant disc flux. This
was done by taking the average disc flux to produce a galaxy template
and compare this to the galaxy component in each image, excluding the
central region. The background level was then fine-tuned to minimize
their difference. By following this procedure we could subtract the
background accurately and fix the disc flux to the template value. The
final nuclear flux was determined by fitting once more the nuclear
component to all the background-corrected images.

\rm
To estimate the error in the determination of the nuclear flux we made
dedicated Monte Carlo simulations using synthetic images. We generated
images with the template-galaxy disc parameters and appropriate PSFs
for the nuclear component. We used 10 different disc-to-nucleus flux
ratios around the measured values. Observational noise was then added
before running the fitting routine on the synthetic data set. We ran
the process for 170 synthetic images to estimate the performance of
the decomposition routine in the specific conditions of the data. The
errors in the determination of the nuclear flux were 3.5\%,
independent of the flux ratio in the range studied. 

\begin{figure}[h]
\psfig{figure=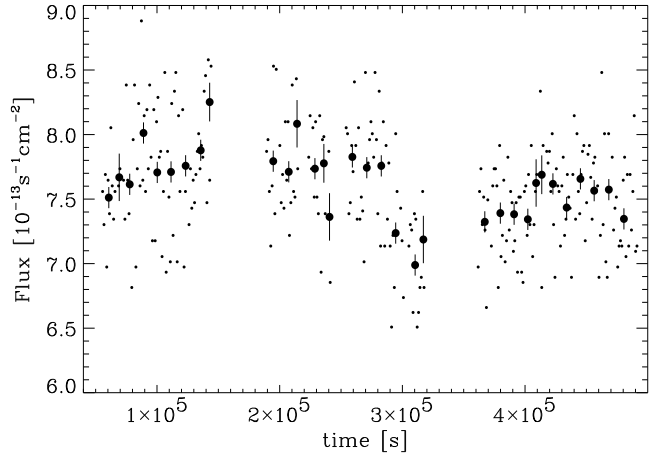,width=8.5cm,height=6.0cm,%
 bbllx=80pt,bblly=440pt,bburx=491pt,bbury=722pt,clip=}
\caption{$U$-band light curve; dots show the unbinned data and filled circles the 10-ks-binned light curve. For clarity, only the error bars of the binned light curve are shown. The errors were calculated using Monte Carlo simulations as described in the text. The time is in seconds, counted from 0:00:00 hrs, July $31^{st}$, 2001.}  
\label{OM_LC} 
\end{figure}

The light curve was constructed by plotting the nuclear flux as a
function of the midpoint time of each exposure and is shown in
Fig. \ref{OM_LC}. To reduce the scatter produced partly by
observational noise we binned the data in 10 ks ($\sim 10$ exposures
per bin) and calculated the errors by averaging quadratically the
errors of the individual points within the corresponding bin. In this
way the intrinsic low amplitude variability of the light curve becomes
evident; it can be seen as the filled circles overlaid on the same
plot. The binned UV light curve varies smoothly on time scales of
$\sim 100$ ks with a maximum amplitude variation of 15\% throughout
the observation, small compared with the variations by a factor of 5
observed in the X-rays.  The average $U$-band nuclear flux measured is
$\sim 7.60\times 10^{-13}$ erg/s/cm$^2$. Assuming a distance of 30 Mpc
(Hayashida et al. \cite{Ha}) and galactic extinction of
$A_{\lambda}=0.334$ (Schlegel et al. \cite{Schle}), the average
nuclear luminosity in the $U$ band is $\sim 1\times 10^{41}$ erg/s.
For comparison we also constructed the light curve of the star used
for PSF modeling using simple aperture photometry. The resulting
10-ks-binned light curve showed small fluctuations, a $\chi ^2$ test
gave a value of $\chi ^2_{red}=1.34$ for 35 degees of freedom against
the constant flux hypothesis. These fluctuations are somewhat larger
than those expected purely from Poissonian noise and could reflect
variations in the sensitivity of the detector. However, given their
small amplitude these variations could only account for a fraction of
the fluctuations seen in the nuclear light curve which gives a value
of $\chi ^2_{\rm red}=6.89$ for the same test.
 
\section{Cross-correlation analysis} 
\begin{figure}[h]
\psfig{figure=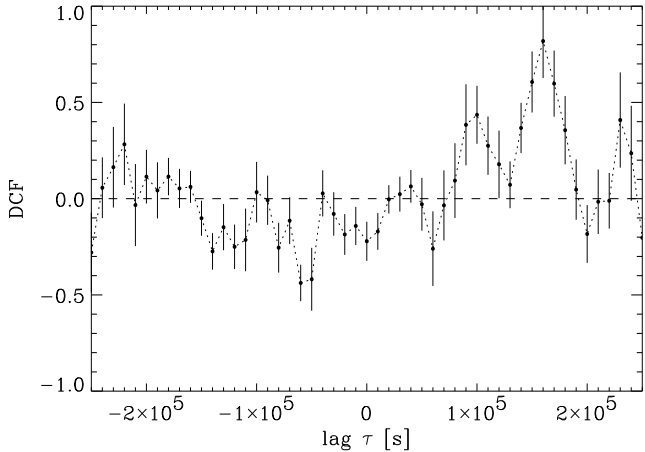,width=8.5cm,height=5.986cm,%
   bbllx=71pt,bblly=439pt,bburx=492pt,bbury=722pt,clip=}
\caption {Discrete Correlation Function (DCF) calculated between the full X-ray band (0.2--10keV) and the UV light curve as a function of time lag in seconds. Positive lags imply UV leading X-rays. The DCF binning used is 10 ks. The error bars plotted represent the standard error of the DCF only and do not account for the effects of the correlations between nearby data points. }
\label{DCF}
\end{figure}

The cross correlation between the
UV and X-ray light curves was computed using the discrete correlation
function (DCF) method of Edelson \& Krolik (\cite{Edel}). The X-ray data were
binned in 5 ks bins to smooth out the rapid variations, which are not
observed in the UV. The UV light curve was binned averaging five
consecutive points to reduce the scatter and the error of the
individual points. The DCF between the full X-ray band (0.2--10 keV)
and the UV light curve using a lag bin of size $\Delta \tau=10$ ks is
plotted in Fig. \ref{DCF}. The most significant feature is
the positive correlation peak at $\tau_{\rm max} = 160$ ks with amplitude
$DCF_{\rm max}=0.82$ where the variations in the UV lead those in the X-rays. No significant peaks, either positive or negative,
are evident in negative lags i.e. there is no indication of UV
variations being driven by changes in the X-ray flux on the time scales probed. 

In order to estimate the error on $\tau_{\rm max}$ and $DCF_{\rm max}$ we
used the Monte Carlo method described by Peterson et al. (\cite{Pe}). We find $\tau=160^{+45}_{-65}$ ks, while
$DCF_{\rm max}=0.82^{+0.17}_{-0.25}$ where the errors represent the 95\%
confidence limits.  These errors are associated mainly with the
uncertainties caused by the observational sampling of the light curves
and the flux uncertainty in the individual measurements. However,
since both light curves are probably realisations of a red-noise
process, it is possible to detect significant correlations even if
they are intrinsically uncorrelated. In order to investigate this
issue further, we performed a second numerical experiment.

As already mentioned in Sect. 2.1, the X-ray power spectrum has a
broken power-law shape. Vaughan et al. (\cite{Vaug1}) found a slope of
$-1$ up to a frequency of $10^{-4}$ Hz above which the slope steepens
to a value of $\sim -2$ for one of their fitted models.  Assuming this
power spectral shape, we used the method of Timmer \& K\"onig
(\cite{Ti}) to construct 10,000 synthetic light curves with a mean and
variance equal to those of the 0.2--10 keV light curve. The points in
the synthetic light curves were sampled every 10 s in order to account
for aliasing effects, while the length was 10 times larger than that
of the observed X-ray light curve in order to account for red-noise
leak. These synthetic light curves were then re-binned in 5 ks bins
and sampled to match the length and sampling pattern of the original
X-ray light curve. We estimated the DCF between the observed UV light
curve and each of the synthetic light curves, registering the
resulting $DCF_{\rm max}$. Only in $1.5\%$ of all cases did we find a
$DCF_{\rm max}$ value larger than 0.82, at any lag. We repeated
this procedure using slightly flatter and steeper power spectrum
slopes with nearly identical results. We conclude that the
positive correlation that we detect in Fig. \ref{DCF} is significant
in the sense that most probably it is due to an intrinsic coupling
between the observed UV and X-ray flux variations. In the same
simulations we found peaks with $|DCF_{\rm max}| > 0.4$ in
approximately 90\% of the cases indicating that the peaks seen at
$\tau \sim$ -60, 100 and 230 ks are not significant.

\begin{figure}[h]
\psfig{figure=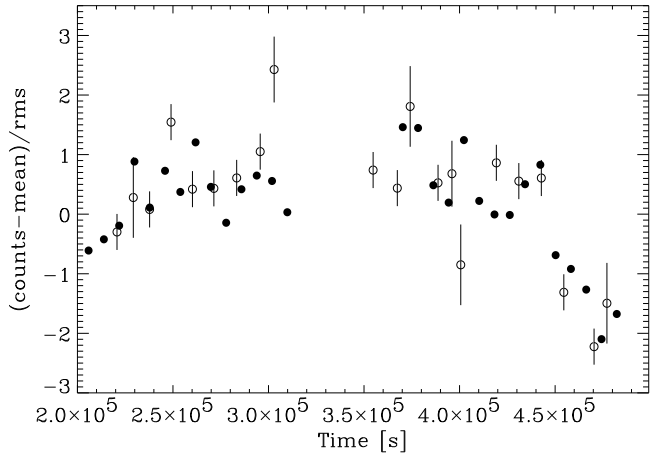,width=8.5cm,height=6.0cm,%
   bbllx=85pt,bblly=440pt,bburx=491pt,bbury=722pt,clip=}
\caption {Overlap region of the normalized UV light curve (open circles) in 10 ks bins and back-shifted X-ray light curve (filled circles) in 8 ks bins. The X-ray light curve was shifted back in time by 160 ks. The X-ray error bars are smaller than the symbols.}
\label{overlay}
\end{figure}

Figure \ref{overlay} shows a plot of the X-ray and UV light curves
normalized to their mean (filled and open circles, respectively).
The X-ray light curve is back-shifted by 160 ks and the plot shows 
the period where the light curves overlap. The light curves are
binned in order to eliminate the fast, large amplitude variations  of the X-ray flux and to
increase the signal-to-noise ratio in the case of the UV. This figure reveals clearly the reason for the large
DCF peak value at lag 160 ks. If we take into account the delay, the
modulation of the UV flux matches well that of the smoothed X-ray light
curve over the interval sampled. The correlation suggests an intrinsic link between the
observed variations in the two energy bands on time scales of $\sim$
10-100 ks, which operates with a delay of $\sim$ 160 ks. 
 
We also investigated briefly the correlation between the UV and the X-ray
light curves in different energy bands. The resulting DCFs are very
similar to that shown in Fig. \ref{DCF}, as is expected since the
different X-ray energy bands show similar variations. In all
cases, we observe a strong peak at lag $\sim 160$ ks, although its
amplitude shows minor differences. We find that the 2--4 keV and 8--10
keV light curves are the best correlated with the UV,
both with a value of $DCF_{\rm max}=0.84$. The 4--7-keV-band light curve,
which is representative of the Fe K$_{\alpha}$ line emission, shows a marginally weaker correlation ($DCF_{\rm max}=0.83$), while the soft-excess-component
(0.2--1.0 keV) light curve shows the weakest correlation with the UV light
curve with a peak of $DCF_{\rm max}=0.78$. 

\subsection{DCF between the UV flux and the variations in the X-ray energy spectrum}

In order to investigate possible correlations between the UV flux and
the variations in the X-ray energy spectrum we used the 2--4 and 8--10 keV
light curve to construct the 2--4 keV/8--10 keV softness ratio
curve. These energy bands were selected to show the spectral
changes in the 2--10 keV continuum while minimizing the contribution of the
Fe line and the soft excess. The softness ratio so defined is
representative of the slope of the continuum, where higher values
correspond to steeper spectra. We used softness ratio and UV light
curves in 5 ks bins and lag bin size of $\Delta \tau=$ 10 ks to
calculate the DCF, and found a moderately good correlation
($DCF_{\rm max}=0.67$) again at a time lag of $\sim$ 160 ks.  We should point out that this value of the $DCF_{\rm max}$ is found in approximately 10\% of the Monte Carlo trials described above and so the correlation is formally not significant on its own.

\begin{figure}[h]
\psfig{figure=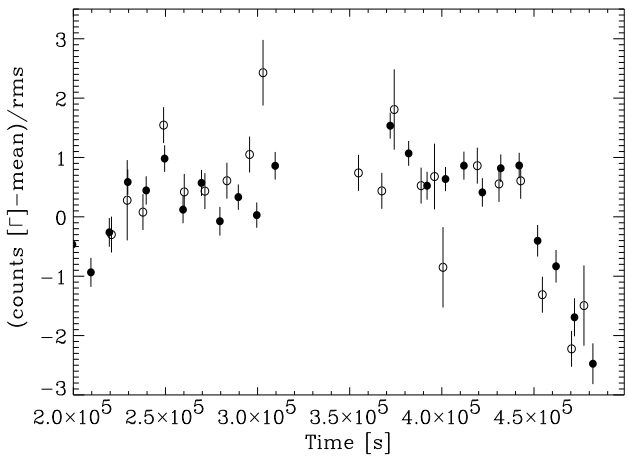,width=8.5cm,height=6cm,%
   bbllx=71pt,bblly=439pt,bburx=492pt,bbury=722pt,clip=}
\caption {$U$-band light curve in 10 ks bins (open circles) overlaid on the photon-index curve of the hard X-ray spectrum fitted with a power-law model (filled circles). The photon-index ($ \Gamma $) curve has been back-shifted by the 160 ks corresponding to the highest peak in the DCF between softness ratio and $U$-band light curves. Both curves were mean-subtracted and divided by their rms.}
\label{gamma}
\end{figure}

The X-ray spectrum in these energy bands can be well fitted by a
power-law model, with photon index $\Gamma \sim 2.0$. To investigate
the behaviour of the spectral slope we fitted a power-law model to the
2--4 and 8--10 keV bands every 10 ks. Fig. \ref{gamma} shows a plot of
the UV light curve (open circles) and of the photon-index (\G) curve
normalized to their mean, where the photon-index curve has been
back-shifted by 160 ks. Like in Fig. \ref{overlay} we show only the
period where the UV flux and the \G\ curves overlap. The figure shows
that the UV flux is well correlated with the X-ray spectral slope, the
2--10 keV continuum flattens (i.e. \G\ decreases) as the UV flux
decreases, with a time delay of 160 ks.

\section{Discussion} The connection between the UV and X-ray emission
from AGN is thought to be primarily due to two processes:
Comptonisation of UV photons into X-rays and reprocessing of X-rays
into the UV range. Here we discuss both mechanisms and confront their
predictions with our results.

\subsection{Comptonisation scenario}

The positive peak found in the DCF with UV variations leading those in the
X-rays suggest that variations in the thermal component (accretion
disc) drive the changes in the emission of the scatterer (corona) on
time scales $\geq 10$ ks, and constitutes evidence for the Comptonisation
scenario.

\subsubsection{Origin of the time lag}

To investigate the possible origin of the time lag we estimate the
physical separation of the emitting regions. The X-ray emission region
is thought to be compact and located very close to the black hole due
to its rapid high-amplitude variability and by the spectral shape of
the fluorescent Fe line. Fabian et al. (\cite{Fab2}) limit the size of
the hard X-ray source to a few $R_{\rm S}$. The $U$-band emission is
probably produced thermally through viscous dissipation in the
accretion disc. We use the standard thin disc model to estimate the
radii of maximum $U$-flux emission. For this we assume a disc radial
temperature profile given by $T_{\rm eff}(R)=[(3GM \dot{M} /8 \pi
\sigma R^{3})(1-\sqrt{R_{\rm in}/R})]^{1/4}$ where $M$ is the black
hole mass, $\dot{M}$ is the accretion rate estimated from the
bolometric luminosity, $\sigma$ is the Stefan-Boltzmann constant, $R$
is the radius in the disc and $R_{\rm in}$ is the radius of the inner
edge of the disc (see eg. Kato et al. \cite{Ka}). We consider the disc
to be composed of rings of approximately uniform temperature radiating
locally as black bodies. For a black hole mass between $10^6$ and
$10^7 M_{\odot}$ (Hayashida et al. \cite{Ha}; Czerny et
al. \cite{Czer}; Vaughan et al. \cite{Vaug1}) and bolometric
luminosity of $7\times 10^{43}$erg/s (Czerny et al. \cite{Czer}) the
rings producing the largest flux in the $U$ band (3000--4000 \AA) are
at $R\sim 100 R_{\rm S}$ for a $10^{6}M_{\odot}$ black hole and $R\sim
25 R_{\rm S}$ for a $10^7 M_{\odot}$ black hole. The radii obtained
correspond to light travel times between the emitting regions of only
a few ks, too short to explain the measured lag.

However, if the accretion disc indeed extends down to a few $R_{\rm
S}$ as the shape of the Fe line suggests, most of the disc flux should
be emitted in this small radius and mainly in higher energy photons
than the observed 3000--4000 \AA~band. Therefore the bulk of the seed
photons for Comptonisation may be emitted in a different region of the
accretion disc than the observed $U$ band photons. The length of the
time lag that we observe favours models where variations in the
accretion flow affect first the flux at outer radii and then in the
innermost region. In this case, the observed delay would correspond to
the time needed for the accretion flow fluctuations to travel inwards
from the radius where most of the $U$ band photons are produced ($\sim
25-100 R_{\rm S}$) to the innermost region of $\sim 5 R_{\rm S}$ where
most of the seed photons for Comptonisation are emitted. As a
reference time scale, the sound crossing time in a standard thin disc
between these radii with black hole mass of $10^6 M_{\odot}$ is $\sim
1000$ ks, marginally consistent with the measured lag. If we consider
that the variable part of the UV light curve might arise from radii
smaller than that producing most of the $U$ band luminosity we obtain
shorter sound crossing times, closer to the 160 ks lag. Alternatively,
MHD effects might transport perturbations through the accretion disc
faster than sound waves and remain a plausible source for the measured
lag.

\subsubsection{Spectral changes}

The Comptonisation scenario is also supported by the correlated
variations that we observe between the UV flux and the slope of the
X-ray continuum energy spectrum. We find that when the UV flux rises
the continuum becomes steeper with a delay of $\sim 160$ ks. This
effect is qualitatively consistent with the behaviour of a hot plasma
receiving more seed photons, radiating more X-rays and cooling to a
new equilibrium temperature, thus producing a steeper power-law
spectrum. A similar UV flux/\G\ correlation has also been observed in
the case of NGC~7469 (Nandra et al. \cite{Nan}).

\begin{figure}[h]
\psfig{figure=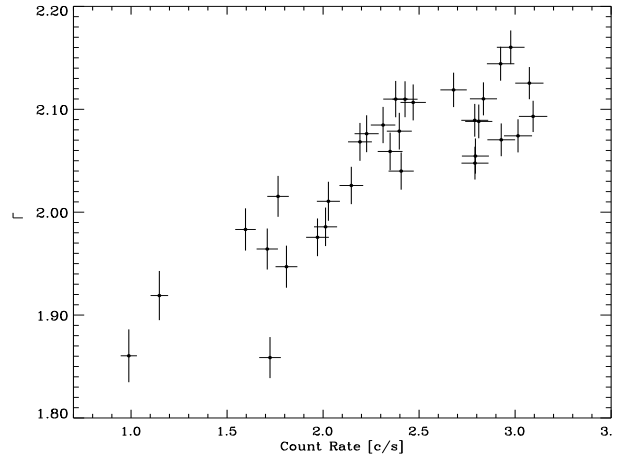,width=8.0cm,height=6cm,%
   bbllx=85pt,bblly=384pt,bburx=491pt,bbury=721pt,clip=}
\caption {Photon index $\Gamma$ of the power-law fitted to the 2--4 keV plus 8--10 keV spectrum as a function of the count rate in these bands.}
\label{HHR}
\end{figure}

We should point out, however, that the UV flux/\G\ correlation in
\mcg\ may be merely a by-product of the UV/X-ray flux variability
correlation. Fig. \ref{HHR} shows a plot of the best-fitting \G\
values as a function of the total count rate in the 2--4 plus 8--10
keV bands.  The photon index $\Gamma$ is well correlated with the
count rate, with higher flux corresponding to a steeper
spectrum. However, intrinsic \G\ variations in \mcg\ have been
questioned in the past using this and other data sets. For example,
Taylor et al. (\cite{Tay}) and Fabian \& Vaughan (\cite{Fab2}) explain
the X-ray spectral variations of the source with the combination of
two components: a constant slope, variable normalization power-law
component and constant reflection component. In this case there are
no genuine \G\ variations, as the softening of the spectrum is due only
to a change in relative normalization of the two fixed shape
components. If this is the case then the UV/\G\ correlation results
merely from the UV/X-ray flux correlation through the good
correspondence between X-ray flux and spectral slope. The situation is
far from clear, though, as the presence of a constant reflection
component may not be able to account for the spectral slope variations
of the source as shown by Papadakis et al. (\cite{Pa}). Since we find
that both the UV/X-ray-flux light curves and the UV flux/\G\ time
series are well correlated with the same time delay, the present
results do not help to discriminate between the two possibilities.

\subsection{Reprocessing scenario}
There is observational evidence that the hard X-ray source in this
object illuminates cooler nearby material, possibly the accretion
disc. The broad and skewed Fe K$_{\alpha}$ line from fluorescent
emission and the reflection hump detected in the X-ray band above 10
keV (Vaughan \& Fabian \cite{Vaug2}) give support to this
statement. It is also expected that the X-rays illuminating such
optically thick material would produce UV emission through thermal
reprocessing and so produce correlated X-ray and UV light curves with
X-ray variations leading the UV´s. However, the DCF between these
light curves showed no positive peaks with negative lags (ie. with
X-rays leading the UV) and so give no indication of reprocessing.
 
To estimate the amount of reprocessed emission expected from the
accretion disc we consider a simple model where the X-rays received by
the disc are totally absorbed and re-emitted as black-body
radiation. We start by calculating the expected $U$-band flux emitted
by a standard thin disc around a $10^6-10^7 M_{\odot} $ black hole
with bolometric disc luminosity $L_{\rm bol}=10^{43.85}$ erg/s (Czerny
et al. \cite{Czer}) through viscous dissipation alone. We obtain the
local black-body spectrum of the disc as a function of radius and then
integrate the flux in the 3000--4000 \AA\ band from $R=3R_{\rm S}$ to
$R=3000R_{\rm S}$. We then recalculate the $U$-band flux now including
the effect of reprocessing. For this simple model we assume that the
reprocessed flux adds to the flux from viscous dissipation:
\begin{displaymath}
F(R)=F_{\rm viscous}(R)+F_{\rm reprocessed}(R)
\end{displaymath}
\begin{displaymath}
= \frac{3GM\dot{M}}{8\pi R^3} \left( 1-\sqrt{\frac{R_{\rm in}}{R}} \right) +\frac{L_{\rm X} ~h}{4\pi(R^2+h^2)^{3/2}}
\end{displaymath}
where $R$ is the radius in the disc and $h$ the height of the X-ray
source over the disc. We take the illuminating X-ray flux $L_{\rm
X}=10^{43}$ erg/s and assume the emitting region to be compact, close
to the disc axis, and at a height $h=3-10 R_{\rm S}$. The integrated
$U$-band flux increases by only 3--12\% when the illuminating X-ray
component is included, where the lower flux results from the smaller
height of the X-ray source. More importantly, a change by a factor of
4 in the X-ray flux induces only a change of 4--11\% in the total
$U$-band flux for the same range of $h$ values. Considering the
simplicity of the model these numbers can only be regarded as
reference values. Moreover, they are only upper limits for this
geometrical configuration, given that here we assume total
reprocessing of the X-ray flux. We conclude that large variations of
the X-ray flux can produce variations in the reprocessed component
which are too small to be detected in this data, especially if the
X-ray source is located at a small height $h$.

\subsection{Comparison with previous results}

Several studies of X-ray/UV/Optical correlations in AGN have been
carried out in the past. From these only the short-term correlations
found can be compared with the work presented here. We have no
information on the correlation between
the optical/UV and X-rays on long (months/years) time
scales for \mcg , and the correlated variability properties of other objects
appear to depend on the time scales probed. Some objects for which
similar studies have been presented are NGC 4051, NGC 7469 and NGC
5548.

Notably Mason et al. (\cite{Ma02}) found the 2900 \AA\ UV lagging the
2-10 keV X-ray continuum by 0.2 days from a 1.5 day long \xmm\
observation of NCG 4051. The mass of this object is similar to that of
\mcg\ and the variability amplitudes found for the X-ray and UV bands
are comparable with those presented here. However they found a lag of
opposite sign and interpret it as evidence for reprocessing, contrary
to what we find in \mcg\ . These results might be reconciled by noting
that NGC 4051 was observed in a higher-frequency UV band. If the
higher-energy UV emission comes from smaller radii of the accretion
disc then this band can be affected more strongly by a compact X-ray
source close to the centre. In this way the UV-reprocessed component,
which is probably present in both objects, would be comparatively
stronger in this observation of NGC 4051. A longer (2 month)
monitoring of this object gave evidence for X-rays lagging the optical
by $\sim$ 2.4 days (Shemmer et al. \cite{She}), similar in sign and
magnitude to the lag we found in \mcg\ between X-ray and $\sim$ 3500
\AA\ bands.

As mentioned before, Nandra et al. (\cite{Nan}) found a good correlation
between UV flux and X-ray spectral index peaking at 0 lag from a 30
day long observation of NGC 7469. This result is similar to what we
observe in \mcg\ considering that a lag of $\sim$ 100 ks would not
be clearly discernible in their daily binned data. In NGC 5548 Chiang
et al. (\cite {Chi}) found the extreme UV (0.1 keV) continuum leading
the variations in the soft and hard X-rays by 10 and 40 ks
respectively from a 20 day baseline campaign. Again this lag is
qualitatively similar to what we find, with lower energies leading the
variations of higher energy bands.

On long time scales however the behaviour of the objects studied can
be quite different. NCG 5548 shows high-amplitude, highly correlated
variability in X-ray and optical bands (Uttley et al. \cite{Uttley}),
NGC 3516 also displays high-amplitude variability in both bands but
this time the correlation is very weak (Maoz et al. \cite{Maoz}) while
NGC 4051 shows very weak optical variability, although well correlated
with the much more variable X-rays (Peterson et al. \cite{Pe00}).

It is possible to explain some of the different X-ray/UV long-term
correlations observed in different objects by considering the
differences in their black-hole masses as noted by Uttley et
al. (\cite{Uttley}). Since \mcg\ shows a similar behaviour to both NGC
5548 ($\sim 10^8 M_{\odot}$) and NGC 4051 ($\sim 10^6 M_{\odot}$) on
short time scales, a longer-term optical campaign would be necessary to
test the optical variability/black-hole-mass correlation in this
object. Apart from this hypothesis it is hard to draw a consistent
picture of the behaviour of all the cases studied given the small
number of objects with similar observations and the large number of
relevant intrinsic parameters. However, we can note that Uttley et
al. (\cite{Uttley}), from long-term optical variability amplitude,
Shemmer et al. (\cite{She}), from considerations of the size of the
X-ray-emitting region, and the work presented here, from the magnitude
of the observed time lag between X-ray and UV variations, all agree
that the connection between these energy bands is probably due to
accretion-flow fluctuations traveling inwards across the emission
regions.

\section{Conclusions}
We analyzed simultaneous UV and X-ray data of \mcg\ to study the relation between the time variability in these two energy bands on time scales of a few ks to a few 100 ks. 
The X-ray light curve displayed large amplitude variations reaching a maximum-to-minimum count rate ratio of $\sim 5$ which implies a change in luminosity of $\Delta L \sim 10^{43}$ erg/s. The UV light curve was calculated using a sophisticated decomposition technique to determine the nuclear flux accurately against the background of the bright host galaxy. This light curve showed much smaller amplitude variability than the X-rays, not larger than 15\% during the whole observation, amounting to $\Delta L < 2 \times 10^{40}$ erg/s. 

We found a significant correlation ($DCF_{\rm max}=0.82$) between the light curves with a time lag $\tau=160^{+45}_{-65}$ ks, where the UV variations lead the X-rays. We interpret the sign of this time lag as evidence for models where X-rays arise from Comptonisation of thermal UV photons, as in an accretion disc-corona scenario. The time lag is too long to represent the light travel time from the $U$-band emitting region of a standard disc to the centre, where the X-rays are thought to be produced. In turn this length favours models where the variations in the disc flux arise from perturbations in the accretion flow traveling inwards through the disc, affecting first the $U$-band emitting region and then the innermost radii from where the bulk of the seed photons for Comptonisation must come. Finally the cross-correlation analysis gave no significant correlation with negative time lags (ie. with X-ray leading UV) suggesting that the bulk of the observed UV does not arise from reprocessed X-rays. 

\begin{acknowledgements}
This work is based on observations with \xmm, an ESA science mission
with instruments and contributions directly funded by ESA Member States
 and the USA (NASA). We are very grateful to H. Spruit and P. Uttley for useful discussions and to M. Freyberg for OM data-processing assistance. We also thank the anonymous referee for useful comments. PA acknowledges support from the International Max Planck Research School on Astrophysics (IMPRS).
\end{acknowledgements}


\begin{thebibliography}{}
\bibitem[2003]{Ball} Ballantyne, D. R., Vaughan, S., \& Fabian, A. C. 2003, MNRAS, 342, 239 
\bibitem[2001]{Czer} Czerny, B., Nikolajuk, M., Piasecki M., \&  Kuraszkiewicz, J. 2001, MNRAS, 325, 865
\bibitem[2000]{Chi} Chiang, J., Reynolds, C. S., Blaes, O. M., et al. 2000, ApJ, 528, 292
\bibitem[1948]{deV} de Vaucouleurs, G. 1948, Annales d'Astrophysique, 11, 247
\bibitem[1988]{Edel} Edelson, R.A., \& Krolik, J.H. 1988, ApJ, 333, 646
\bibitem[2002]{Fab1} Fabian, A. C., Vaughan, S., Nandra, K., et al. 2002, MNRAS, 335, L1
\bibitem[2003]{Fab2} Fabian, A. C., \& Vaughan, S. 2003, MNRAS, 340, L28 
\bibitem[1970]{Free} Freeman, K. C. 1970, ApJ, 161, 802
\bibitem[2001]{Ghi} Ghizzardi, S., \& Molendi, S. 2001, Proc. of the conference 'New Visions of the X-ray Universe', ESTEC Nov. 2001
\bibitem[1988]{Gui} Guilbert, P.W., \& Rees, M.J. 1988, MNRAS, 233, 475
\bibitem[1991]{Haardt} Haardt, F., \& Maraschi, L., 1991, ApJ, 380, 51
\bibitem[1998]{Ha} Hayashida, K., Miyamoto, S., Kitamoto, S., Negoro, H., \& Inoue, H. 1998, ApJ, 500, 642
\bibitem[1996]{Iwa} Iwasawa, K., Fabian, A. C., Reynolds, C. S., et al. 1996, MNRAS, 282, 1038
\bibitem[1998]{Ka} Kato, S., Fukue, J., Mineshige, S. 1998, Black-Hole Accretion Disks, Kyoto (Kyoto University Press) 
\bibitem[2004]{Kuhl} Kuhlbrodt, B., Wisotzki, L., \& Jahnke, K. 2004, MNRAS, 349, 1027
\bibitem[2002]{Maoz} Maoz, D., Markowitz, A., Edelson, R., \& Nandra, K., 2002, AJ, 124, 1988
\bibitem[2002]{Ma02} Mason, K.O., McHardy I.M., Page M. J. et al. 2002, ApJ, 580, 117  
\bibitem[2003]{Ma} Mason, K.O., Breeveld, A., Much, R. et al. 2001, A\&A, 365, 36
\bibitem[2003]{Mi}  Miniutti, G., Fabian, A. C., Goyder, R., \& Lasenby, A. N. 2003, MNRAS, 344, 22
\bibitem[2000]{Nan} Nandra K., Le, T., George, I.M., Edelson, R.A., et al. 2000, ApJ, 544, 734
\bibitem[2002]{Pa} Papadakis, I. E., Petrucci, P. O., Maraschi, L., et al. 2002, ApJ, 573, 92
\bibitem[1998]{Pe} Peterson, B. M., Wanders, I., Horne, K., et al. 1998, PASP, 110, 660
\bibitem[2000]{Pe00} Peterson, B. M., McHardy, I. M., Wilkes, B. J., et al. 2000, ApJ, 542, 161
\bibitem[1998]{Schle} Schlegel, D., Finkbeiner, D., \& Davis, M., 1998, ApJ, 500, 525
\bibitem[2003]{She} Shemmer, O., Uttley, P., Netzer, H., McHardy, I.M. 2003, MNRAS, 343, 1341
\bibitem[2001]{Strue} Str\"uder, L.,  Briel, U.G., Dennerl, K.,  et al. 2001, A\&A, 365, 18
\bibitem[1980]{Sun} Sunyaev, R., \& Titarchuck, L.G. 1980, A\&A, 86, 121
\bibitem[1995]{Ta} Tanaka, Y., Nandra, K., Fabian, A. C., et al. 1995, Nature, 375, 659
\bibitem[2003]{Tay} Taylor, R., Uttley, P., \& McHardy, I.M. 2003, MNRAS, 342, 31
\bibitem[1995]{Ti} Timmer, J., \& K\"onig, M. 1995, A\&A, 300, 707
\bibitem[2003]{Tur1} Turner, A. K., Fabian, A. C., Vaughan S., \& Lee, J. C. 2003, MNRAS, 346, 833
\bibitem[2004]{Tur2} Turner, A. K., Fabian, A.C., Lee, J., \& Vaughan, S. 2004, astro-ph/0405570
\bibitem[2003]{Uttley} Uttley, P., Edelson, R., McHardy, I. M., et al. 2003, ApJ, 584, 53
\bibitem[2003]{Vaug1} Vaughan, S., Fabian, A.C., \& Nandra, K. 2003, MNRAS, 339, 1237
\bibitem[2004]{Vaug2} Vaughan, S., \& Fabian, A. C. 2004, MNRAS, 348, 1415
\end{thebibliography}
\end{document}